\documentclass[draft,jgrga]{agutex}

  \usepackage{graphicx}
  \usepackage{subscript}
  \usepackage{lmodern}
  \usepackage{float}
\authorrunninghead{KRAUSS ET AL.}

\titlerunninghead{Thermosphere and geomagnetic response to ICMEs}

\authoraddr{Corresponding author: S. Krauss,
Space Research Institute, Austrian Academy of Sciences, Schmiedlstrasse 6, A-8042 Graz, Austria
(sandro.krauss@oeaw.ac.at)}

\begin{document}

%
%
\title{Thermosphere and geomagnetic response to interplanetary coronal mass ejections observed by ACE and GRACE: Statistical results.}
%
%
\authors{S. Krauss,\altaffilmark{1} M. Temmer,\altaffilmark{2}
A. Veronig,\altaffilmark{2} O. Baur\altaffilmark{1} and H. Lammer\altaffilmark{1}}

\altaffiltext{1}{Space Research Institute, Austrian Academy of Sciences, Schmiedlstr. 6, A-8042 Graz, Austria}
\altaffiltext{2}{Institute of Physics, University of Graz, Universit{\"a}tsplatz 5/II, A-8010 Graz, Austria.}

%
%
\begin{abstract}
For the period July 2003 to August 2010, the interplanetary coronal mass ejection (ICME) catalogue maintained by Richardson and Cane lists 106 Earth-directed events, which have been measured in-situ by plasma and field instruments on--board the ACE satellite.  We present a statistical investigation of the Earth's thermospheric neutral density response by means of accelerometer measurements collected by the GRACE satellites, which are available for 104 ICMEs in the data set, and its relation to various geomagnetic indices and characteristic ICME parameters such as the impact speed ($v_{\rm max}$), southward magnetic field strength ($B_{\rm z}$). The majority of ICMEs causes a distinct density enhancement in the thermosphere, with up to a factor of eight compared to the pre--event level. We find high correlations between ICME $B_{\rm z}$ and thermospheric density enhancements ($\approx 0.9$), while the correlation with the ICME impact speed is somewhat smaller ($\approx 0.7$). The geomagnetic indices revealing the highest correlations are Dst and SYM-H ($\approx 0.9$), the lowest correlations are obtained for $k_{\rm p}$ and AE ($\approx 0.7$), which show a nonlinear relation with the thermospheric density enhancements. Separating the response for the shock sheath region and the magnetic structure of the ICME, we find that the Dst and SYM-H reveal a tighter relation to the $B_{\rm z}$ minimum in the magnetic structure of the ICME, whereas the polar cap indices show higher correlations with the $B_{\rm z}$ minimum in the shock sheath region.
Since the strength of the $B_{\rm z}$ component---either in the sheath or the magnetic structure of the ICME---is highly correlated ($\approx 0.9$) with the neutral density enhancement, we discuss the possibility of satellite orbital decay estimates based on magnetic field measurements at L1, i.e. before the ICME hits the Earth magnetosphere.
These results are expected to further stimulate progress in space weather understanding and applications regarding satellite operations.
\end{abstract}

%
%
\begin{article}

\section{Introduction}

Since the beginning of the space age in the late 1950's, ``space weather'' is a hot topic. It deals with the space environment of the Earth's atmosphere from approximately 90\,km altitude onwards extending all the way to the Sun. Strong disturbances of this space environment are primarily caused by solar flares, coronal mass ejections (CMEs) and associated solar energetic particles (SEPs). Owing to advances in solar research in the recent decades, today we know that CMEs cause the most comprehensive spectrum of space weather disturbances, including geomagnetic storms, aurorae, geomagnetically induced currents affecting electric power lines, heating of the outer atmosphere, and many more.

\begin{figure}[htb]
 \begin{center}
  \includegraphics[width=8cm]{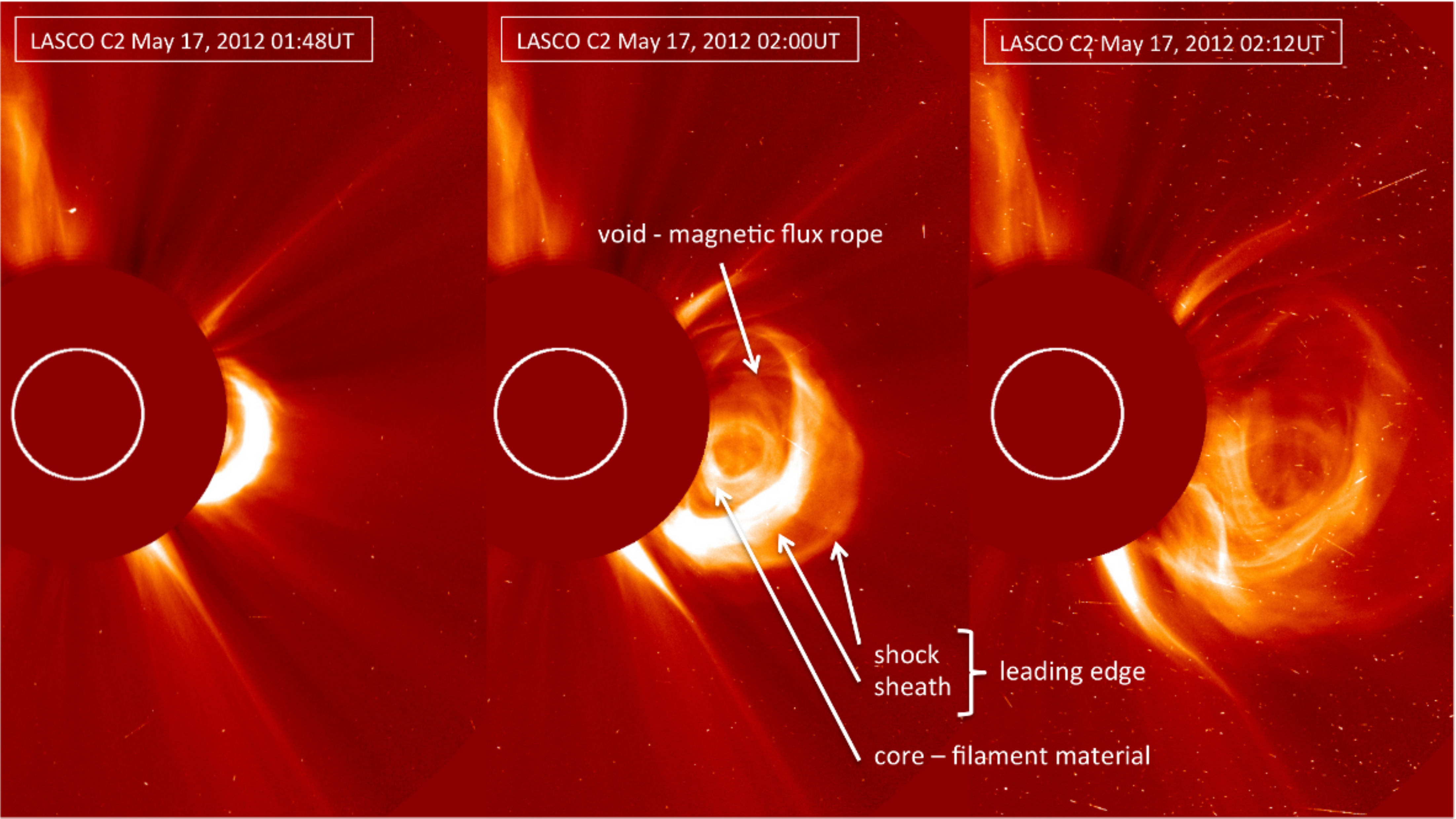}
  \vspace{1ex}
 \end{center}
 \caption{Illustration of the CME three-part structure. Sequence of SOHO/LASCO C2 images of a CME that occurred on May 17, 2012 close to the Western limb.}
 \label{fig:1}
\end{figure}

CMEs are huge clouds of magnetized plasma propagating from the solar corona into interplanetary space with speeds ranging from a few hundred up to $\approx~$3000 km~s$^{-1}$. CMEs often reveal a three-part structure \citep{Illing1985} (Fig.~\ref{fig:1}): the front (leading edge) with its shock-sheath region \citep{Vourlidas2013}, the void which comprises the magnetic structure of a CME (often with a magnetic flux rope inherited) and the bright core consisting of filament material. The occurrence rate of CMEs is closely related to the solar cycle, and varies from 0.3 per day at solar minimum to 4--5 per day on average at solar maximum \citep[e.g.,][]{StCyr2000,Gopalswamy2006}.

Close to the Sun, the CME dynamics is governed by the driving Lorentz forces resulting from magnetic instabilities in the fields of active regions \citep{Zhang2001,Vrsnak2007,Bein2012}.

In interplanetary space, on the other hand, the CME dynamics is governed by the interaction of the CME with the surrounding solar wind flow, whereby fast CMEs are decelerated and slow CMEs are accelerated \citep[e.g.,][]{Gopalswamy2000}. This interaction process with the collisionless plasma in the interplanetary space is described by the magnetohydrodynamic equivalent of the aerodynamic drag force \citep{Cargill2004,Vrsnak2013}. Thus, the transit time and the arrival speed of an Earth-directed CME---which are both crucial parameters for space weather description and forecasting---depend on both the initial speed and the state and interaction of the ambient solar wind plasma with the CME. Typical transit times range between 2--5 days, depending on initial speed, mass and size as well as the speed and density of the surrounding solar wind plasma. The fastest CME events observed so far reached the Earth in less than one day \citep[e.g.,][]{Liu2014,Temmer2015}.

Part of the solar wind energy supplied to the Earth's magnetosphere is continuously deposited into the thermosphere via particle precipitation and Joule heating.
This energy supply can be enormously enhanced when a fast CME with a strong southward magnetic field component arrives at the Earth's atmosphere, causing a strong geomagnetic storm \citep{Gosling1991}.
The enhanced energy input from the solar wind to the magnetosphere and its ultimate dissipation in the thermosphere causes heating and expansion of the Earth's thermosphere \citep{Krauss2012}. During periods of high CME activity, short timescale (1--2 days) variations in the thermosphere are driven by magnetospheric energy input rather than by changes in the solar extreme-ultraviolet (EUV) flux \citep[e.g.,][]{Wilson2006,Guo2010,Krauss2012}. \citet{Knipp2004} have shown that in case of extreme events the geomagnetic power contributes two-thirds of the total solar power at the Earth's thermosphere, with the major contribution from Joule heating.

The increase in the neutral density affects Earth-orbiting satellites in such a way that the drag force on the spacecraft is enhanced. The fact that some low-Earth orbiters (LEOs) are equipped with accelerometers allows the derivation of in-situ measurements of the current state of the Earth's thermospheric density. To this class of satellites belong the spacecraft of the projects CHAllening Mini satellite Payload \citep[CHAMP;][]{Reigber2002} and Gravity Recovery And Climate Experiment \citep[GRACE;][]{Tapley2004}.

In this paper, we present a statistical study of the thermospheric response to a sample of 104 interplanetary CMEs (ICMEs); most of these events caused a significant increase in the neutral density. The period under investigation covers seven years during the maximum and decaying phase of solar cycle 23 (7/2003 to 8/2010), including the ``Halloween storms''. Thermospheric density changes are derived from GRACE accelerometer measurements and compared with ICME properties measured by the Advanced Composition Explorer \cite[ACE;][]{Stone1998} and various geomagnetic storm indices. Due to the high correlations obtained, we were able to derive regression curve parameters, which we expect to be highly valuable for the improved estimates of future CME-related thermospheric density enhancements.

\section{Data and Analysis}
For the present study we analyzed all ICME events during the time range July 13, 2003 until August 04, 2010 as listed in the catalogue of near-Earth interplanetary ICMEs maintained by Richardson \& Cane (R\&C) \citep{Richardson2003,Richardson2010}. During this period the R\&C list contains 106 ICME events, comprising the arrival times of the shock-sheath region, the start and end times of the magnetic structure (based on the plasma and magnetic field observations) and the minimum value of the geomagnetic disturbance storm time index (Dst) \citep{Sugiura1964,SugiuraKamei1991} within the total ICME duration. For our statistical study we thoroughly analyzed the magnetic field characteristics of the ICMEs making use of measurements by the ACE satellite located at the Lagrange point L1. The response of the Earth's thermosphere in terms of changes in neutral density are derived from acceleration measurements provided by the GRACE mission, which has been orbiting the Earth at an altitude of $\sim$450-500\,km since 2002. Furthermore, we used various geomagnetic indices derived from terrestrial magnetic field measurements collected at ground-based observatories.

\subsection{Magnetic field measurements}
The ICME kinematics and its associated magnetic field were derived from in-situ measurements recorded by instruments aboard the ACE spacecraft -- in particular the Solar Wind Electron, Proton, and Alpha Monitor (SWEPAM) \citep{McComas1998} and the magnetometer (MAG) \citep{Smith1998}. We used level-2 data of the proton speed (solar wind bulk speed) and the magnetic field ($B _{\rm z}$ component in Geocentric Solar Ecliptic, GSE, coordinates) with a time resolution of 4\,min (OMNI database) \citep{King2004}. In GSE coordinates the $z$-component is oriented orthogonal to the ecliptic, from which we derived the southward component of the magnetic field. We identified minimum values in the $B _{\rm z}$ component separately for the shock-sheath region and the magnetic structure of the ICME. Bad or missing data were linearly interpolated. Time information, such as start and end of the disturbance, were extracted from the R\&C list.

\begin{figure}[htb]
 \begin{center}
  \includegraphics[width=8cm]{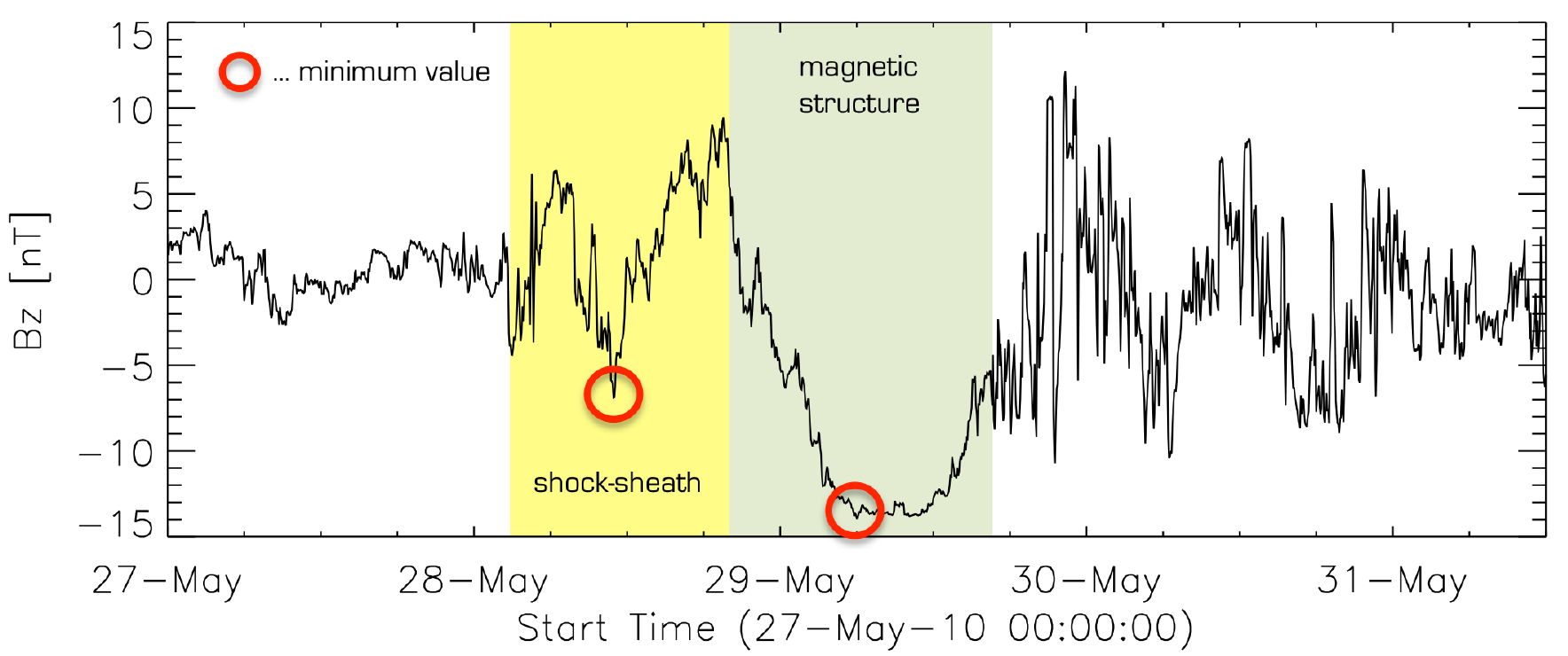}
  \vspace{1ex}
  \end{center}
  \caption{$B_{\rm z}$ observations from ACE during an ICME event in May 2010. Red circles indicate the minimum values used in the further analysis. Yellow and green areas mark the shock-sheath region and magnetic structure, respectively, according to the time information provided in the R\&C list.}
 \label{fig:2}
\end{figure}

Figure~\ref{fig:2} shows the temporal evolution of the $B _{\rm z}$ component, exemplary for an ICME event in May 2010. The shock-sheath region covers a period of strongly varying magnetic field (Fig.~\ref{fig:2}, yellow area), with a minimum occurring on May 28 at approximately 11:00~UT. The second minimum occurs about 19 hours later when the magnetic structure of the ICME arrived (Fig.~\ref{fig:2}, green area). During this period the $B _{\rm z}$ component shows a smooth behavior and changing orientation, indicative of field rotation. The existence of a shock wave depends on the speed of the ICME relative to the ambient solar wind speed. For 82 of the ICME events analyzed in this study, a shock-sheath region could be identified in the in-situ plasma and field data.

\subsection{Thermospheric neutral density}
In order to gain knowledge about the response of the Earth's thermosphere to ICME events we analyzed accelerometer data collected by the US-German GRACE project. The GRACE satellites were launched in March 2002 and injected into a low-altitude (initially 505\,km), near-circular (eccentricity $\approx~0.001$), and near-polar (inclination $\approx~89^{\circ}$) orbit around the Earth. The GRACE constellation consists of two identical spacecraft following each other in the same orbit, separated by about 220\,km.

GRACE is a dedicated gravity field mission, and therefore both satellites are equipped with accelerometers; the GRACE accelerometers were manufactured by the Office National d'Etudes et de R\'{e}cherches Aerospatials (ONERA) and have an accuracy of $\approx~1\times~10^{-10}$m~s$^{-2}$. As far as gravity field recovery is concerned, accelerometer measurements are used to get rid of the non-gravitational forces acting on the satellites. Here, we use these observations to compute neutral densities ($\rho$) along the trajectory of the spacecraft. The transition from the along track accelerometer measurements ($a_{\rm x}$) to atmospheric neutral densities can be established by

\begin{equation}
\rho = - \frac{2 a_{\rm x} m}{C_{\rm D} A v_{\rm x}^2},
\label{eq:rho}
\end{equation}

where $m$ indicates the satellite mass, $C_{\rm D}$ the highly complex and variable drag coefficient, $A$ the effective cross sectional area and $v_{\rm x}$ the satellite velocity in the along track axis. For a detailed description of the technique we refer the reader to \citet{Sutton2008,Doornbos2009,Krauss2013}. Due to ellipticity, the true GRACE orbit implies satellite altitude variations of about 50\,km per revolution. For this reason, all densities shown in this study are normalized to an average altitude of 490\,km ($\rho_{\rm 490}$) using the empirical thermosphere model by \citet{Bowman2008} ($\rho_{\rm JB}$, $\rho_{\rm JB490}$) according to

\begin{equation}
 \rho_{\rm 490}= \frac{ \rho_{\rm JB490}}{ \rho_{\rm JB}} ~ \rho.
 \label{eq:norm}
\end{equation}

We smoothed the density time series with a 5-min moving average filter to avoid single peaks biasing the results.

\begin{figure}[htb]
 \begin{center}
  \includegraphics[width=8cm]{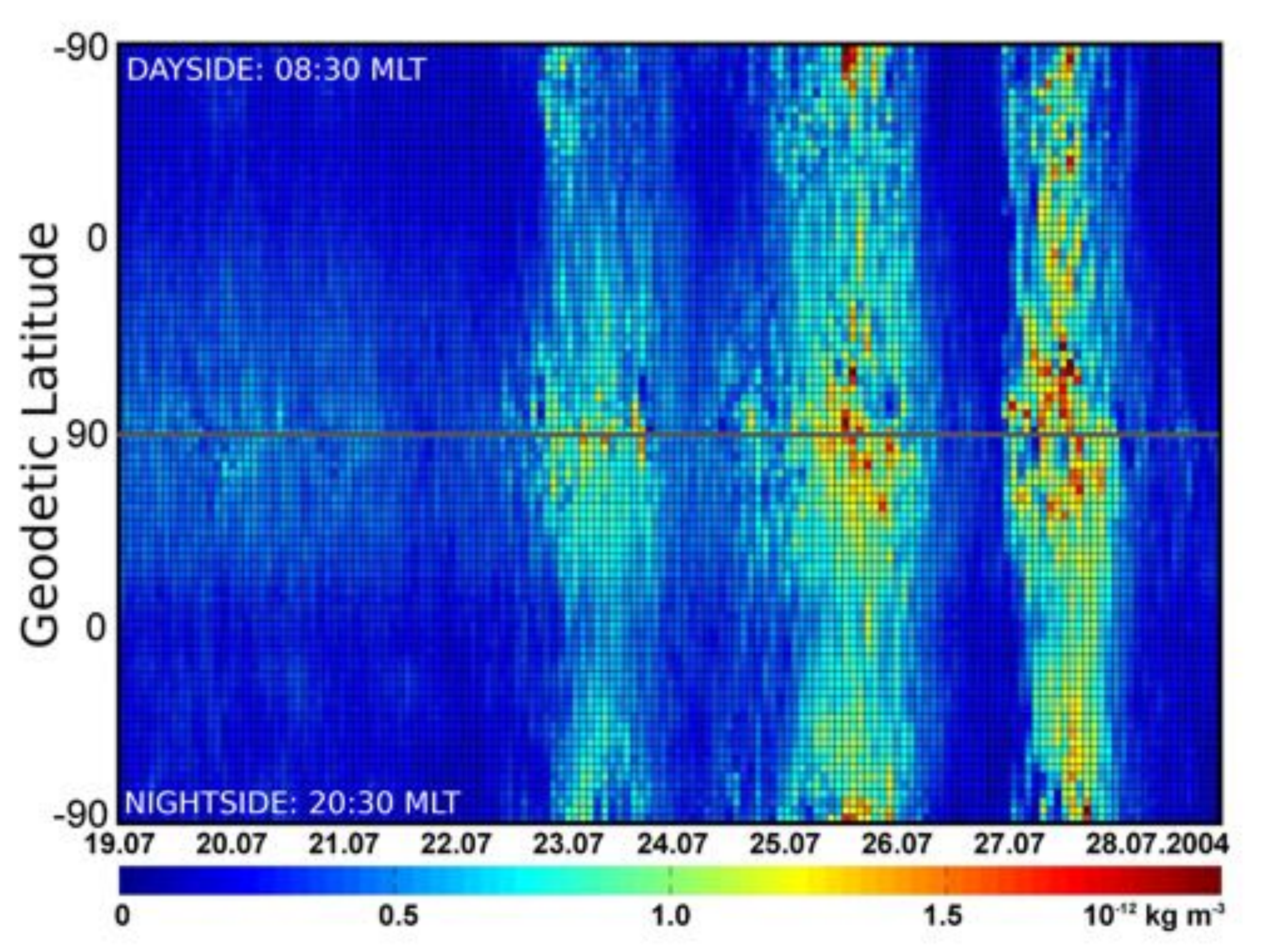}
  \vspace{1ex}
 \end{center}
 \caption{Latitude-time plot of thermosphere neutral densities derived from GRACE satellite accelerations. The impact of three ICMEs in July 2004 inducing severe geomagnetic storms can be identified in the neutral density measurements. The perturbations are visible on both the day--side (top:~08:30 mean local time) and night--side (bottom:~20:30 mean local time). The time interval between two adjacent vertical grid lines covers one satellite revolution (94\,min).}
 \label{fig:3}
\end{figure}

As an example, Fig.~\ref{fig:3} depicts neutral densities in a latitude-time plot covering a period of ten days in July 2004; during this period three ICMEs hit the Earth. Compared to the pre--event level of $\approx~0.3\times10^{-12}$~kg\,m$^{-3}$, each ICME increased the neutral density by a factor of up to five. The perturbations started at high latitudes and thereafter propagated towards the equator within a few hours. These phenomena are known as travelling atmospheric disturbances and occur at high latitudes due to sudden energy and momentum deposition, mostly in the form of Joule heating \citep[e.g.,][]{Vasyliunas2005,Bruinsma2008}. For most of the analyzed ICME events, we found the maximum geomagnetic disturbance at high latitudes near the auroral region (with a latitude root mean square value of $73.2^{\circ}$).

\subsection{Geomagnetic indices}
\begin{figure}[htb]
 \begin{center}
  \includegraphics[width=8cm]{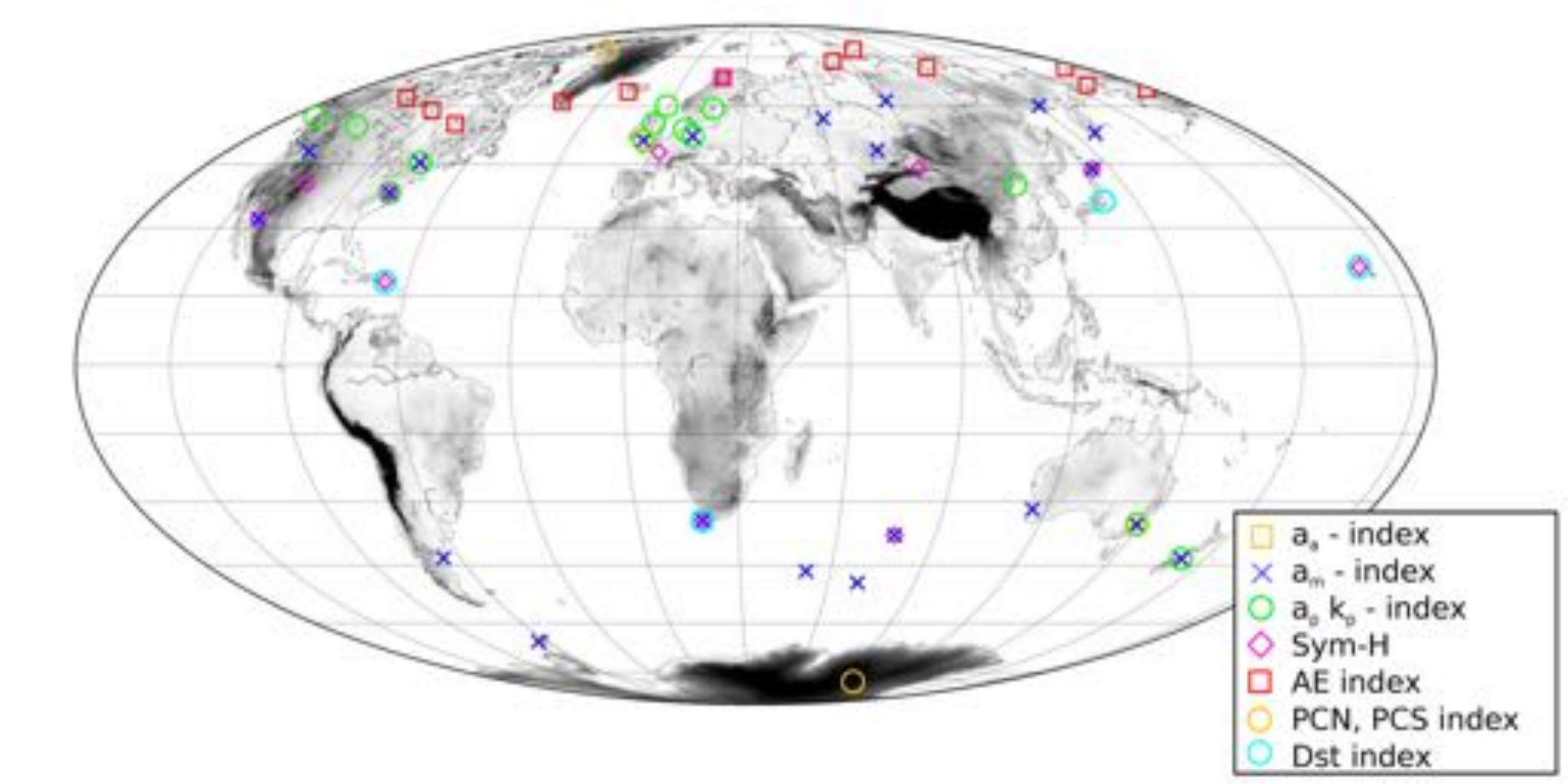}
 \end{center}
 \caption{Location of observatories that provide measurements to construct geomagnetic indices. The indices are provided with different temporal resolution -- a\textsubscript{a}, a\textsubscript{m}, a\textsubscript{p} and k\textsubscript{p}: 3\,h; PC: 1\,min; Dst: 1\,h, AE: 1\,min; SYM--H: 1\,min.}
 \label{fig:4}
\end{figure}

Geomagnetic indices play an essential role in the space weather system, describing the response of the Earth's magnetosphere to effects of the changing solar wind. In the course of the past 80 years various indices were developed, which are monitoring the geomagnetic activity on different latitudes and time scales. Figure ~\ref{fig:4} gives an overview of the global distribution of geomagnetic observatories, which provide the measurements for the different geomagnetic indices.

For this study we analyzed ten different geomagnetic indices. Four of them (a\textsubscript{a} \citep{Mayaud1971}, a\textsubscript{m} \citep{Mayaud1968} a\textsubscript{p} \citep{BartelsVeldkamp1954} and k\textsubscript{p} \citep{Bartels1949}) are deduced from the quasi logarithmic K-index \citep{Bartels1939}---a measure for the variations in the two horizontal field components. The hourly Dst index (derived from observations made by four near-equatorial observatories) describes the global symmetrical geomagnetic perturbations of the ring current---a current flowing westward around the Earth.

Further we analyzed the auroral electrojet index (AE) \citep{DavisSugiura1966}, which is based on observations in the auroral zone and monitors the horizontal field in the east--west ionospheric current; it is constructed from the AU and AL indices, which represent the largest and lowest deviation from a base value at each station, respectively ($\mathrm{AE=AU-AL}$). The polar cap index (PC) \citep{Troshichev1988} represents geomagnetic disturbances in the polar cap regions due to ionospheric and field--aligned currents \citep{Vassiliadis1996}; measurements are performed by the stations Thule and Vostok, therefore the index can be divided into a northern (PCN) and southern (PCS) part, respectively.

Finally, we included the SYM--H index which describes the longitudinal symmetric disturbance field in mid-latitudes for the horizontal component. Basically it is the same as the hourly Dst index and describes the same axial symmetric disturbances in the horizontal component \citep{Iyemori2010}. The main difference is in the temporal resolution of the indices (1h versus 1min), i.e. short-term changes in solar wind parameters are reflected  differently in the two indices. For more information concerning geomagnetic indices the reader is referred to \citet{Menvielle2011}.

\subsection{Analysis}
\begin{figure*}[htb]
 \begin{center}
  \includegraphics[width=15.5cm]{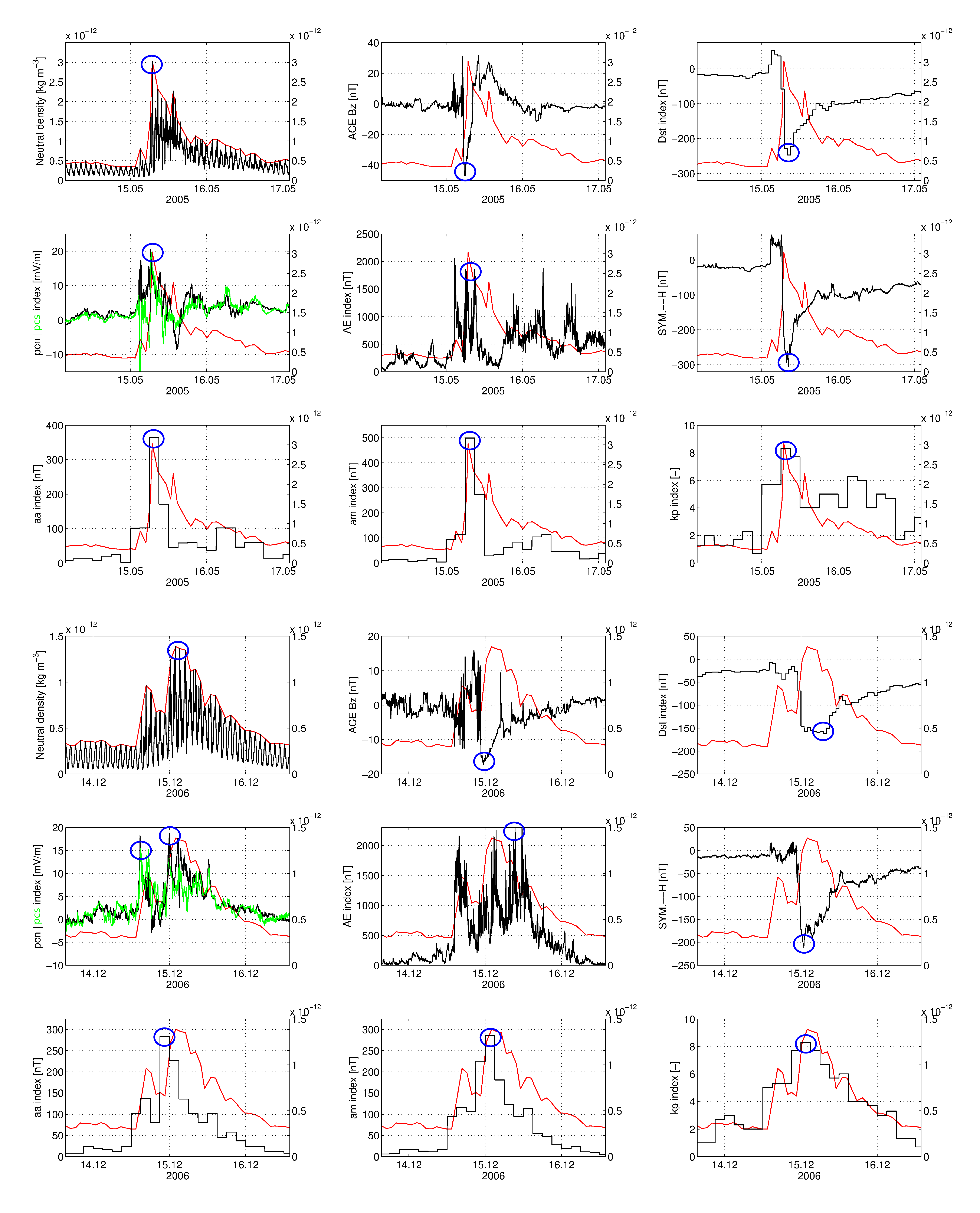}
 \end{center}
 \caption{Overview of analyzed parameters during an ICME event on May 15 in 2005 (rows~a--c) and December 14 in 2006 (rows~d--f). Polar cap index (b1,e1) is split into PCN (black) and PCS (green) part; the red curves shows the evolution of the neutral density; blue circles indicated the peak values of the specific parameters independent whether the peak occurred in the shock-sheath region or the magnetic structure of the ICME.}
 \label{fig:5}
\end{figure*}

The time period July 2003 to August 2010 was chosen because of the availability of GRACE calibration data \citep{Bruinsma2007}. During this time span the R\&C list contains 106 ICME events. For two of them, no GRACE acceleration measurements are available. The remaining set of 104 ICMEs is studied in this paper. Starting with the disturbance time of the ICME (taken from the R\&C list) we defined a time window of 36 hours over which we extracted the evolution of the various geomagnetic indices, the neutral densities (from GRACE) and the magnetic field measurements (from ACE). For every single ICME event, we derived from each parameter the peak value within the predefined time window. We thoroughly checked the data to ensure that the correct peak value is assigned to each ICME event. This is particularly important for time intervals containing more than one ICME event as well as for complex ICME structures.\\
To account for variations in the background thermospheric density (caused by the non-disturbed solar activity) we calculated the absolute density increase, i.e. the difference $\Delta \rho_{\rm 490}$ between the peak density measured during the event minus the pre--event level, as well as the relative density increase in percent; for the correlation studies the $\Delta \rho_{\rm 490}$ values were used. Exemplarily for two ICME events, Fig.~\ref{fig:5} shows the evolution of the individual parameters. Additionally, each panel includes the progress of the neutral density in terms of an envelope function (red curves) as well as the detected peak values (blue circles). The envelope function was created such that it comprises the maximum density per satellite revolution (Fig.~\ref{fig:5}, a1, d1). 

Figure~\ref{fig:5} reveals that the temporal evolution of both the geomagnetic indices and the $B_{\rm z}$ component of the magnetic field is in good agreement with the neutral density. As far as the Dst index and the SYM--H index (a3,b3; d3,e3) is concerned, the different phases of the geomagnetic storm become clearly visible: a sudden decrease of the indices during the main phase of the storm, followed by a much slower recovery phase. Regarding the absolute values, over the entire range we derive slightly higher values in the SYM--H index than in the Dst index and in some cases a small offset between these two indices (d3, e3). This behaviour can be explained by the different methods in the baseline preparations for the two indices \citep{Wanliss2006}.

\begin{figure}[htb]
  \begin{center}
   \includegraphics[width=8cm]{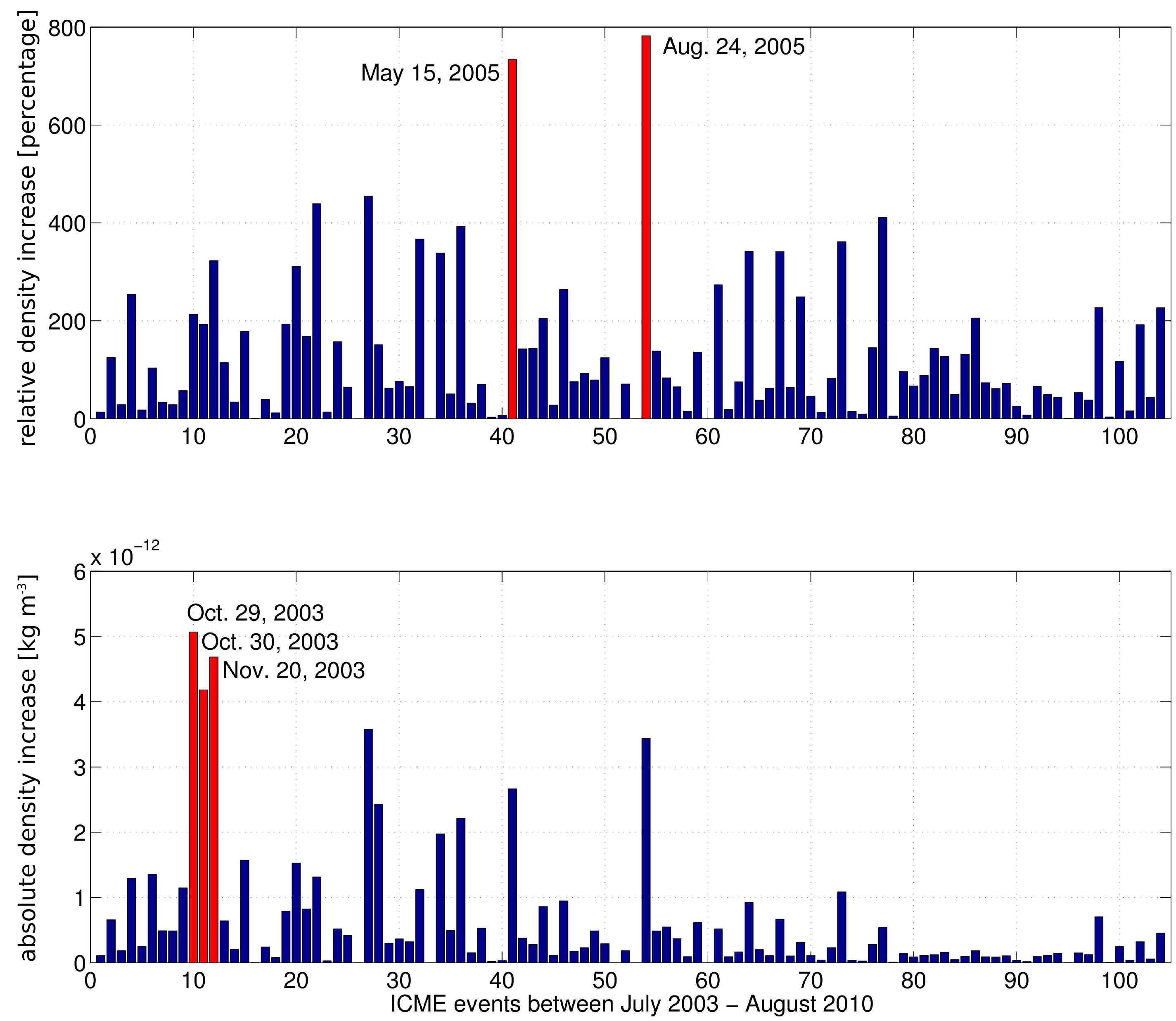}
   \vspace{1ex}
  \end{center}
  \caption{Absolute (bottom) and relative (top) neutral density increase (w.r.t. to pre--event level) for each of the analyzed ICME events. Marked in red are the most extreme events in both cases.}
  \label{fig:6}
 \end{figure}

Notice that the a--indices (row c,f) exhibit a rather rough structure, caused by their low temporal resolution (3h). The AE index shows disturbances throughout the whole ICME events and even beyond that~(b2). Comparing the two polar cap indices (b1,e1) we see larger differences during the most disturbed period than before and after the event, implying different conditions in the polar cap regions \citep{Lukianova2002}.\\
In general, the majority of the analyzed ICME events shows a significant density increase in the atmosphere compared to the density level before the perturbation (Fig.~\ref{fig:6}). The largest relative increase occurred for two events in 2005 (May 15 and August 24); in both cases the density rose by about 750\% compared to the pre--event level. The highest absolute increases in neutral density occurred during the Halloween storms in 2003 (October 29--30 and November 20) with $\Delta \rho_{\rm 490} > 4\times10^{-12}$kg~m$^{-3}$.

\section{Results}
\begin{figure*}[htb]
 \begin{center}
  \includegraphics[width=13cm]{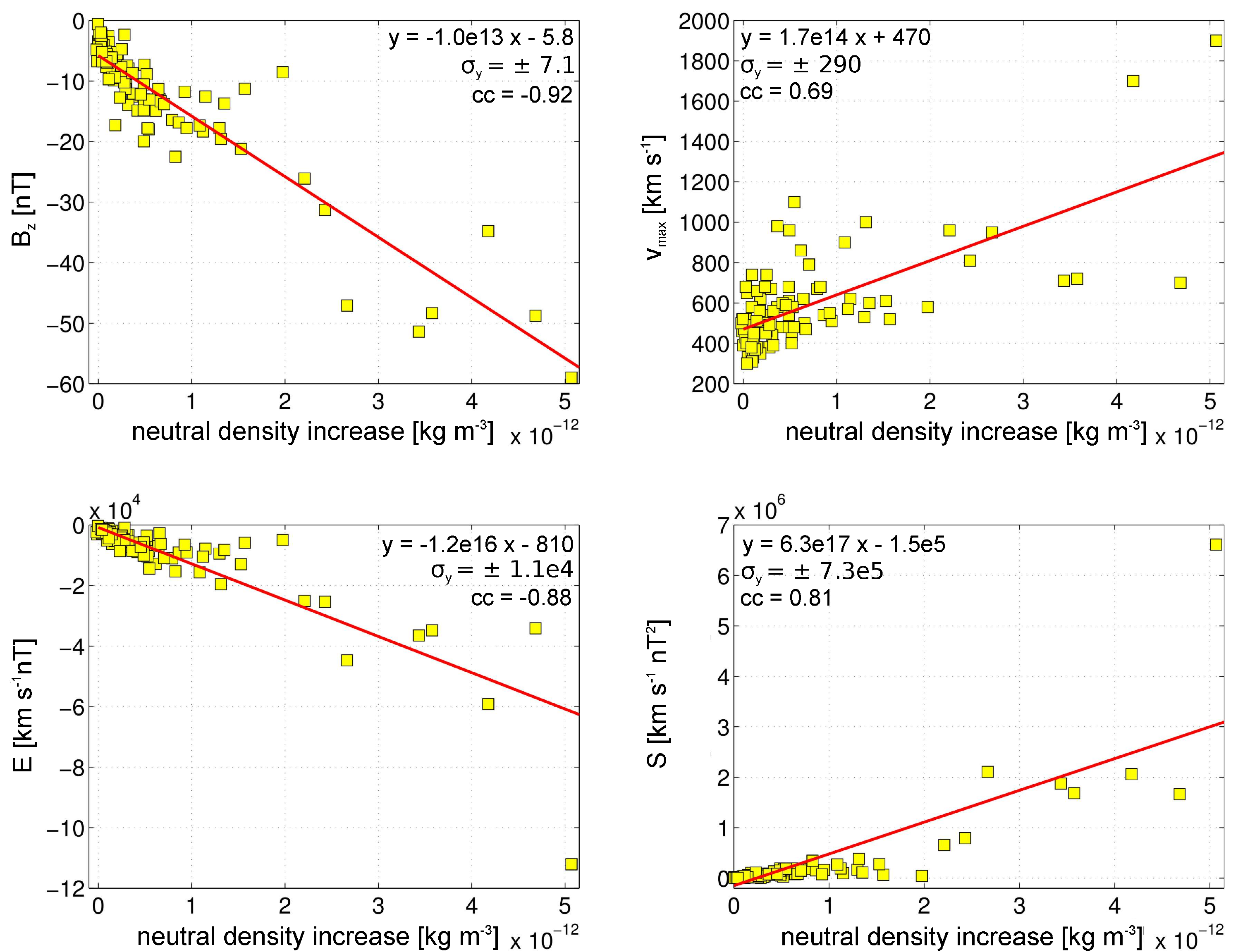}
 \end{center}
  \caption{Scatter plot of thermospheric densities $\Delta \rho_{\rm 490}$ and ICME parameters in terms of $B_{\rm z}$ component in GSE, the maximum ICME velocity ($v_{\rm max}$), the convective electric field estimate $E$ and the Poynting flux estimate $S$. The correlation coefficients and the linear regression coefficients together with the standard deviation for a 90\% confidence interval are given in the inset.}
 \label{fig:7}
\end{figure*}

In a first step, we compare the in-situ measurements by the ACE and GRACE spacecraft in terms of magnetic field strength ($B_{\rm z}$ in GSE) and thermospheric neutral density ($\Delta \rho_{\rm 490}$), respectively. Additionally, we use of the maximum speed ($v_{\rm max}$), measured over the entire ICME structure as given in the R\&C list.\\
Figure \ref{fig:7} shows scatter plots of the thermospheric density increase $\Delta \rho_{\rm 490}$ versus $B_{\rm z}$, $v_{\rm max}$, as well as two combinations of these quantities. On the one hand a proxy for the convective electric field \citep{Burton1975} represented by
\begin{equation}
 E \approx v_{max} \times B_{\rm z},
 \label{eq:E}
\end{equation}
on the other hand a proxy for the energy input via the Poynting flux into the magnetosphere by magnetic reconnection \citep{Akasofu1981} defined as
\begin{equation}
 S = E \times B\approx v_{\rm max}\times B^2_{\rm z}.
 \label{eq:S}
\end{equation}

From the regression analysis we derive the lowest correlation coefficient ($cc=0.69$) for the maximum solar wind speed during the disturbance period ($v_{\rm max}$), inferring that this is not the primary factor for effective driving of geomagnetic storms. In fact, very high impact velocities ($>$ 1500 km~s$^{-1}$) are only present for two of the 104 events. Irrespective of whether a density response occurred or not, the majority of ICMEs has velocities in the range between 400 and 800~km~s$^{-1}$ which is of the statistical average \citep[e.g.,][]{Yurchyshyn2005}.\\
In contrast to $v_{\rm max}$, the quantities $B_{\rm z}$, $E$ and $S$ show a much higher correlation with the thermospheric density with $cc=-0.92$, $cc=-0.88$ and $cc=-0.81$, respectively. The latter correlation coefficient is particularly influenced by one extreme ICME event; excluding the gigantic Halloween ICME (October 29, 2003) from the calculations leads to a significantly increased correlation coefficient between the $\Delta \rho_{\rm 490}$ and $S$ of $cc=-0.90$. Fig.~\ref{fig:7} (top left) also indicates that ICME events with hardly any thermospheric response are associated with a small $B_{\rm z}$ component ($<$10~nT).\\

\begin{figure*}[htb]
 \begin{center}
  \includegraphics[width=15cm]{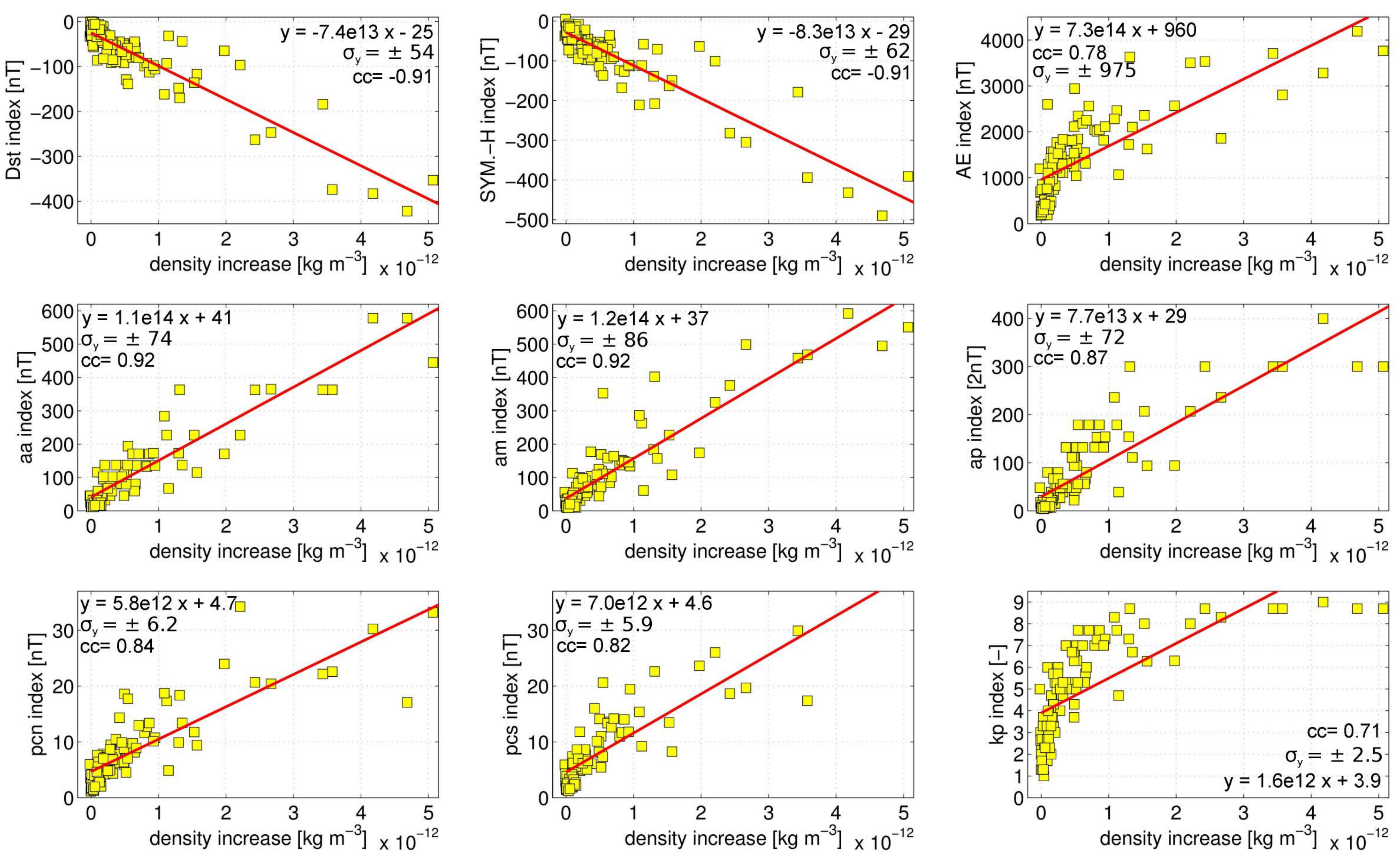}
  \vspace{1ex}
 \end{center}
 \caption{Scatter plot of the increase in the neutral thermospheric density and the various geomagnetic indices. The correlation coefficients and the linear regression coefficients together with the standard deviation for a 90\% confidence interval are given in the inset.}
 \label{fig:8}
\end{figure*}

In the next step we analyzed the correlation between the neutral density increase $\Delta \rho_{\rm 490}$ and the different geomagnetic indices. Figure \ref{fig:8} provides an overview in terms of scatter plots. It can clearly be seen that most of the geomagnetic indices show an excellent correlation with the thermospheric density response.

The highest correlations are derived for the indices $a_{\rm a}$~($cc=0.92$), $a_{\rm m}$~($cc=0.92$) and Dst~($cc=-0.91$) together with the SYM--H index~($cc=-0.91$). The reason for the slightly lower correlation of the $a_{\rm p}$ index might be due to the worse station distribution (13 in the northern hemisphere, 2 in the southern hemisphere). Significantly lower correlations are present for the $k_{\rm p}$ ($cc=0.71$) and AE index ($cc=0.77$). In the first case, this behaviour is not surprising since it is well known that the $k_{\rm p}$ index is not linearly correlated with the geomagnetic activity \citep{BartelsVeldkamp1954}; for this reason the planetary $a_{\rm p}$ index was established in 1954.

Between the density increase and the AE index a non-linear relationship can be observed. Analyzing Fig.~\ref{fig:8} (top, right) we see different behaviours depending on the geomagnetic activity level. Only minor density increases are visible for AE values up to 1000 nT. Above this value, we recognize larger scattering and some indication of saturation. The lower correlation between the AE index and the thermospheric densities is in agreement with previous findings \citep{Akasofu1983,Baumjohann1986,Chen2014}. These authors revealed some limitations concerning the measured quantity (H--field), the existence of longitudinal gaps and the small latitudinal range of the contributing observatories. Compared with the other indices the correlations between the density increases and the polar cap indices (PCN, PCS) are smaller. However, it should be noted that for some ICME events no PCS data was available which excludes some of the strongest events (``Halloween storms'' in October/November 2003).

\begin{table}[ht]
 \caption{Correlation coefficient between the various parameters and the increase in neutral density $\Delta \rho_{\rm490}$ (column 2) as well as the $B_{\rm z}$ component measured by ACE (column 3).}
 \begin{tabular}{lrr} \hline
  {} & {\bf{$\Delta \rho_{\rm 490}$}} & {\bf{$B_{\rm z}$}} \\
  $\Delta \rho_{\rm 490}$  & 1.00 & -0.92 \\
  $B_{\rm z}$  & -0.92 & 1.00 \\
  $v_{\rm max}$  & 0.69 & -0.63 \\
  $E$ & -0.88 & 0.87 \\
  $S$ & -0.81 & 0.81 \\
  Dst index  & -0.91 & 0.88 \\
  AE index  & 0.78 & -0.77 \\
  a\textsubscript{a} index  & 0.92 & -0.88 \\
  a\textsubscript{m} index  & 0.92 & -0.90 \\
  a\textsubscript{p} index  & 0.87 & -0.85 \\
  k\textsubscript{p} index  & 0.71 & -0.74 \\
  PCN index  & 0.84 & -0.79 \\
  PCS index  & 0.82 & -0.76 \\
  SYM--H index & -0.91 &  0.88 \\ \hline
 \end{tabular}
 \label{tab:correlation}
\end{table}

Table~\ref{tab:correlation} summarizes all determined correlation coefficients with regard to the atmospheric density $\Delta \rho_{\rm490}$ and magnetic field component $B_{\rm z}$. In general, the correlations between $B_{\rm z}$ and the geomagnetic indices are very similar to those of the neutral density. Again the highest correlations are found for the $a_{\rm a}$, $a_{\rm m}$, Dst and the SYM--H index. Likewise, the more exponential relation with the AE and $k_{\rm p}$ indices and the slightly poorer correlations with the polar cap indices.\\

\begin{figure*}[htb]
 \begin{center}
  \includegraphics[width=13cm]{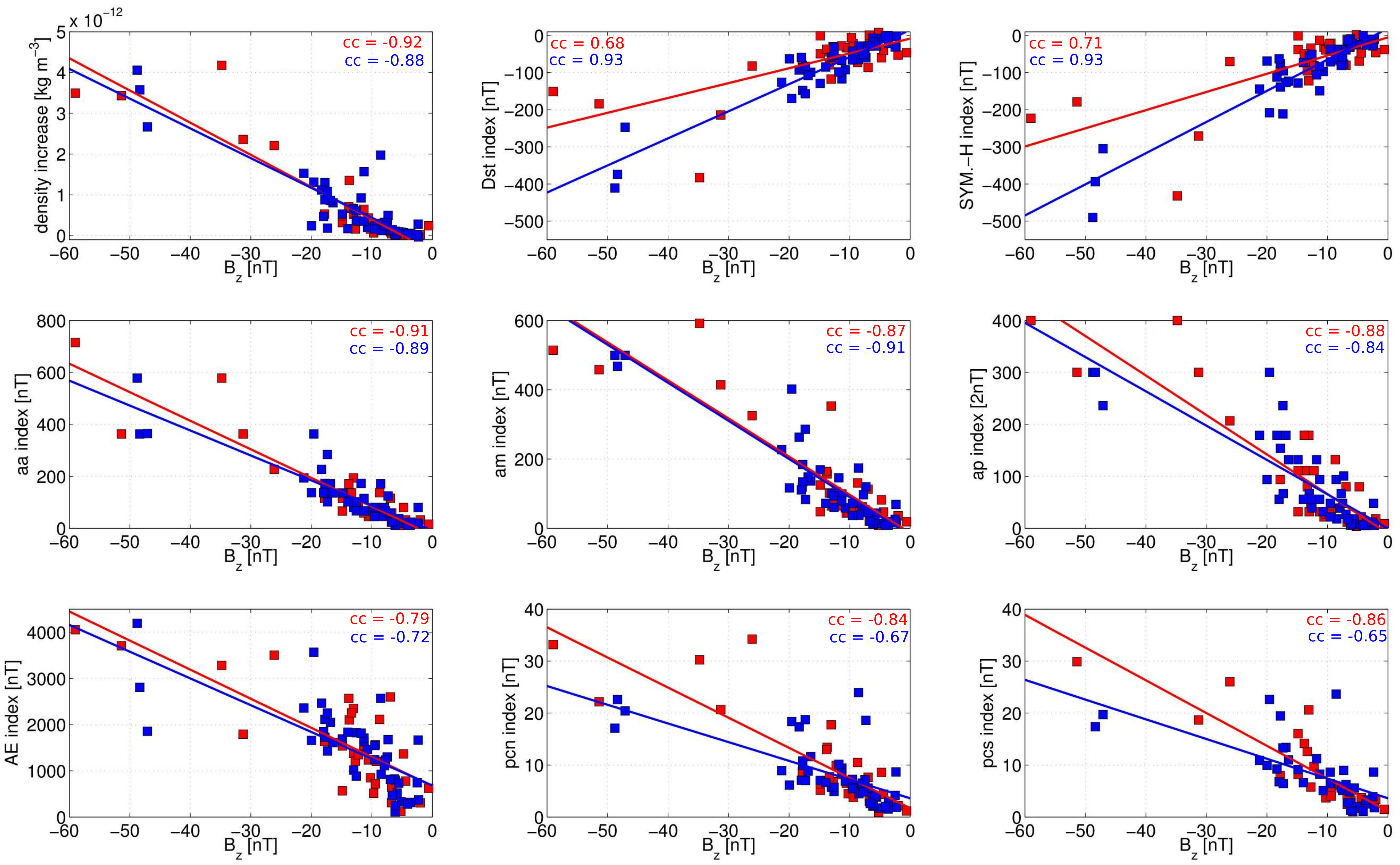}
 \end{center}
\caption{Scatter plot of the minimum $B_{\rm z}$ component as measured by ACE with the absolute density increase and the various geomagnetic indices. Marked in red are those events where the minimum in the $B_{\rm z}$ component is found in the shock-sheath region; blue indicates ICMEs with the $B_{\rm z}$ minimum found in the magnetic structure. The correlation coefficients and the linear regression coefficients are given in the inset.}
 \label{fig:9}
\end{figure*}

Analyzing the minimum in the $B_{\rm z}$ separately for the shock-sheath and the magnetic structure of the ICME we may deduce some additional information. For some events, no shocks are reported, and the leading edge of the ICME might be stated as the estimated arrival time of the disturbance. See \citet{Richardson2003} for additional information on the methods used to identify ICMEs. For 82 out of the 104 analyzed events it was possible to detect and distinguish between the shock-sheath region and the magnetic structure of the specific ICME. For this subset of events we identified the peak values in the neutral density and the various geomagnetic indices separately for the shock sheath region and the magnetic structure of the ICME. As a result, Fig. \ref{fig:9} illustrates the minimum $B_{\rm z}$ component, either found in the shock-sheath region (red squares) or in the magnetic structure part (blue squares) of the ICME, against the neutral density increase and the different geomagnetic indices, for the particular region.

For the a--indices and the AE index no significant differences in the regression line between the two specific regions are visible. However, large differences are present for the polar cap indices (PCN, PCS), indicating that both correlate better with disturbances triggered at the time of arrival of the shock sheath region (red regression line). In contrast to that, the hourly Dst index as well as the SYM--H index show better correlations when the minimum $B_{\rm z}$ component is detected in the magnetic structure of the ICME.
Finally, we examine the time of occurrence of the neutral density and the the peak for the geomagnetic indices. We find the peak times coincide significantly better with the times extracted for the minimum $B_{\rm z}$ value in the magnetic structure rather than for the $B_{\rm z}$ value in the shock-sheath region (ratio: 4:1).\\

The complete list of the selected peak values of all quantities discussed in this study is available as supporting information.

\section{Discussion}
Based on the ICME catalogue maintained by R\&C we have analyzed 104 ICME events for the period July 2003 to August 2010. We studied the relationship between geomagnetic disturbances and thermospheric densities derived from GRACE accelerometer measurements. During periods of geomagnetic disturbances the maximum of the signal was mainly found at high latitudes in the auroral zone (see, e.g., case study by \citet{Bruinsma2006}). By correlating magnetic field characteristics of the ICMEs observed by the ACE satellite we were able to derive linear regression parameters between the different measured quantities. We found that the majority of these ICMEs, cause a significant increase in the Earth's thermospheric neutral density. From the regression analysis, it turned out that the ICME parameter $v_{\rm max}$ has a significantly lower correlation with the thermospheric density than the $B_{\rm z}$ component or the combined parameters $E$ and $S$, which are directly related to the energy input into the system. It is worth noting that the highest correlation coefficient is obtained for $B_{\rm z}$ ($cc \approx 0.9$), whereas the estimates for $E$ and $S$ do not yield improved correlations. Consequently, it is conceivable to obtain an estimate of the forthcoming maximum neutral density increase using in-situ observations of ICME characteristics by appropriate in-situ measurements from satellites at the Lagrange point L1 before the ICME impacts Earth's magnetosphere. Using real-time in-situ data, this would result in a lead time of some tens of minutes (depending on the actual speed of the ICME event). Furthermore, we can conform previous findings by \citet{Liu2010}, that all ICMEs causing a significantly density enhancement have shown a strong negative $B_{\rm z}$ component.

Regarding the various geomagnetic indices we obtained high correlations with the thermospheric density as well as the $B_{\rm z}$ component; especially the SYM--H index, hourly Dst index, $a_{\rm a}$ index and $a_{\rm m}$ index showed a good accordance ($cc \approx 0.9$). However, it is worth noting that, the a--indices have a lower temporal resolution of three hours, which might lead to limitations in detailed space weather monitoring \citep{Menvielle2011}. No linear relationship with the geomagnetic activity could be established for the AE and $k_{\rm p}$ index.

Subsequent investigations concentrated on the separation of the magnetic field observations $B_{\rm z}$ in the shock--sheath region and magnetic structure of the ICME. We found that the hourly Dst index and the SYM--H index correlate better with the $B_{\rm z}$ component when the peak values are extracted from the magnetic structure part of the ICME. Contrary to that, the polar cap indices (PCN, PCS) showed higher correlations with the $B_{\rm z}$ minimum located in shock-sheath region. From a temporal view the occurrence times of the peak values in the neutral density and the geomagnetic indices coincide better (75\% of the cases) with the time of the minimum $B_{\rm z}$ component in the magnetic structure of the ICME. However, strong perturbations from the shock-sheath region may cause an additional enhancement in the atmospheric neutral density~\citep{Yue2011}.

\begin{figure}[htb]
  \begin{center}
   \includegraphics[width=6.5cm]{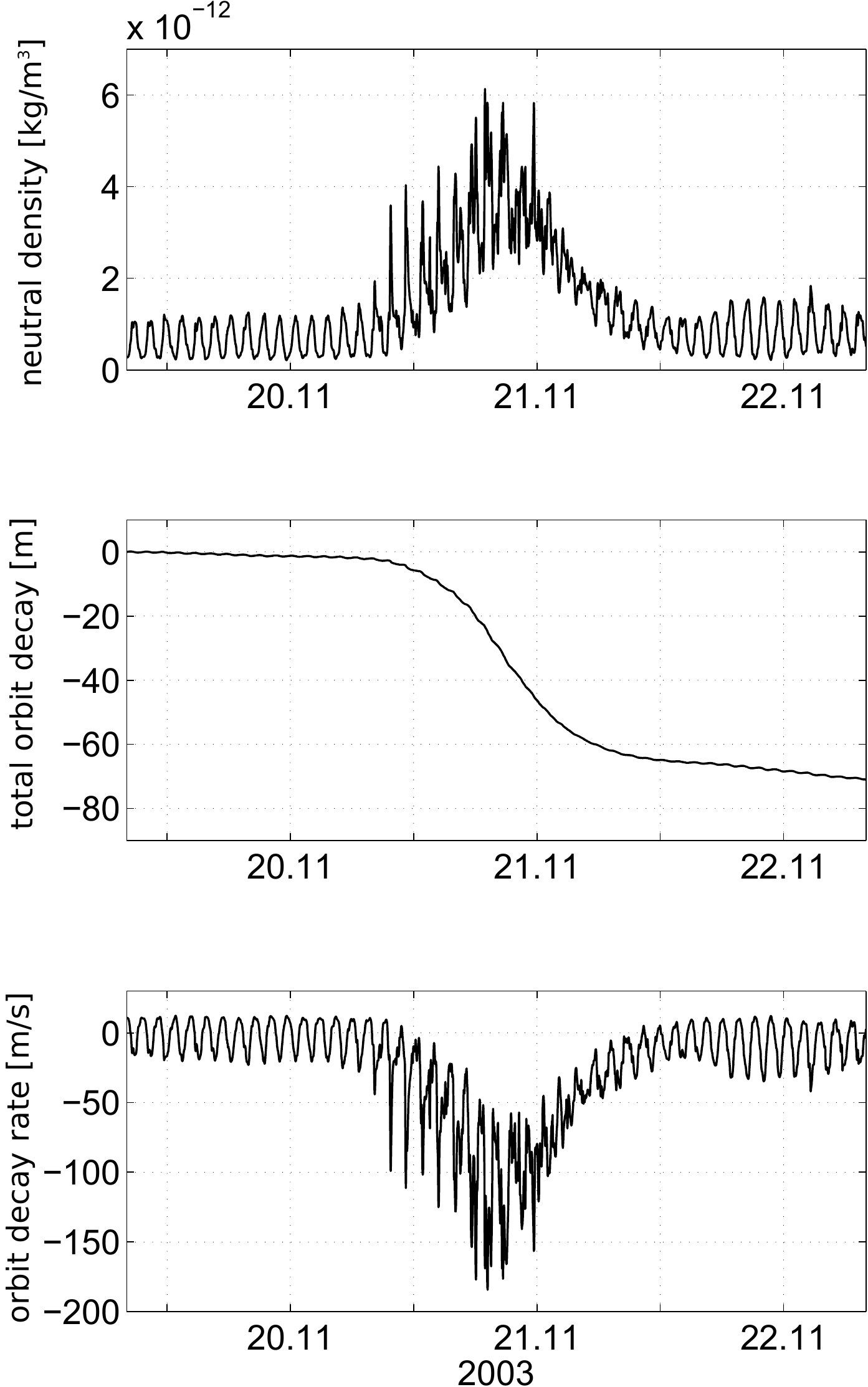}
   \vspace{1ex}
  \end{center}
  \caption{From top to bottom: Atmospheric neutral density increase, total orbit decay and daily orbital decay rate during an ICME event in November 2003.}
  \label{fig:10}
 \end{figure}

Finally, an important field of application of space weather is the induced satellite orbit decay due to variations in the atmospheric density. By using different parameter types ($C_D$, $A$ $\Delta \rho$) elaborated in the present study we determined the orbit decay rate in terms of changes in the mean semi--major axis following \citet{Chen2012}. Figure \ref{fig:10} shows exemplarily, for the ICME event on November 20 in 2003, variations of the GRACE semi--major axis. The middle panel illustrates the total orbit decay over the entire event, whereas the bottom panel shows the orbit decay rate per day.

Comparing the evolution of the neutral density (Fig.~\ref{fig:10}, top) and the orbit decay rate per day (Fig.~\ref{fig:10}, bottom) it becomes apparent that these quantities reveal an almost one-to-one relation. As a result, the distinct linear relationships derived between thermal density enhancements and the ICME $B_{\rm z}$ component can in principle be used to predict the orbital decay rates from the ICME measurements at L1 upstream of Earth. An evaluation of this approach will be the topic of a forthcoming study. Supporting data are included as a table in an SI file; alternatively the data may be obtained from S.K. (email: sandro.krauss[at]oeaw.ac.at).

%

\begin{acknowledgments}
M.T. and A.V gratefully acknowledge the support from the Austrian Science Fund: project nos. FWF V195-N16 and FWF P27292-N20. H.L. acknowledges support from the FWF NFN subproject S11607-N16 'Particle/Radiative Interactions with Upper Atmospheres of Planetary Bodies Under Extreme Stellar Conditions'. S.K. thanks Sean Bruinsma for his support with the GRACE accelerometer calibration.
\end{acknowledgments}


%
%
\end{article}


\begin{thebibliography}{}
\bibitem[Akasofu(1981)]{Akasofu1981}
Akasofu, S.I.,
Energy coupling between the solar wind and the magnetosphere.
Space Science Reviews, 28, 121-190, 1981.

\bibitem[Akasofu et al.(1983)]{Akasofu1983}
Akasofu, S.I., Ahn, B.H., Kamide, Y., Allen, J.H.,
A note on the accuracy of the auroral electrojet indices,
J. Geophys. Res., 88, A7, 5769-5772, 1983.

\bibitem[Bartels et al.(1939)]{Bartels1939}
Bartels, J., Heck, N.H., Johnston, H.F.,
The three -hour-range index measuring geomagnetic activity. J. Geophys. Res., 44 (4), 411-454, 1939.

\bibitem[Bartels(1949)]{Bartels1949}
Bartels, J.,
The standardized index, Ks, and the planetary index, Kp.
IATME Bulletin, 12B, 97, 1949.

\bibitem[Bartels and Veldkamp(1954)]{BartelsVeldkamp1954}
Bartels, J. and Veldkamp, J.
International data on magnetic disturbances, fourth quarter, 1953.
J. Geophys Res, 59, 295, 1954.

\bibitem[Baumjohann(1986)]{Baumjohann1986}
Merits and limitations of the use of geomagnetic indices in solar wind-magnetosphere coupling studies.
Solar Wind--Magnetosphere Coupling, edited by Y. Kamida nad J.A. Slavin, 3--15, Terra Sci., Tokyo, 1986.

\bibitem[Bein et al.(2012)]{Bein2012}
Bein, B.M., Berkebile-Stoiser, S., Veronig, A.M., Temmer, M., Muhr, N., Kienreich, I., Utz, D., Vr\v{s}nak, B.
Impulsive Acceleration of Coronal Mass Ejections. I. Statistics and Coronal Mass Ejection Source Region Characteristics.
The Astrophysical Journal, 738, 191, doi: 10.1088/0004-637X/738/2/191, 2012.

\bibitem[Bowman et al.(2008)]{Bowman2008}
Bowman, B.R., Tobiska, W.K., Marcos, F.A., Huang, C.Y., Lin, C.S., Burke, W.J.,
A New Empirical Thermospheric Density Model JB2008 Using New Solar and Geomagnetic Indices, AIAA/AAS Astrodynamics Specialist Conference, AIAA 2008-6438, 2008b.

\bibitem[Bruinsma et al.(2006)]{Bruinsma2006}
Bruinsma, S., Forbes, J.M., Nerem, R.S., Zhang, X.
Thermosphere density response to the 20--21 November 2003 solar and geomagnetic storm from CHAMP and GRACE accelerometer data. J. Geophys. Res., 111, 1--14, 2006.

\bibitem[Bruinsma et al.(2007)]{Bruinsma2007}
Bruinsma, S., Biancale, R., Perosanz, F.,
Calibration parameters of the CHAMP and GRACE accelerometers.
IUGG XXIV General Assembly, Perugia, 2007.

\bibitem[Bruinsma and Forbes(2008)]{Bruinsma2008}
Bruinsma, S., Forbes, J.M.,
Properties of traveling atmospheric disturbances (TADs) inferred from CHAMP accelerometer observations. Advances in Space Research, 3, Issue 3, p. 369-376, 2008.

\bibitem[Burton et al.(1975)]{Burton1975}
Burton, R.K., McPherron, R.L., Russell, C.T.,
An Empirical Relationship Between Interplanetary Conditions and Dst.
J. Geophys. Res., 80 (31), 4204-4214, 1975.

\bibitem[Cargill(2004)]{Cargill2004}
Cargill, P.~J., On the Aerodynamic Drag Force Acting on Interplanetary Coronal Mass Ejections. Solar Physics, 221, 135-149,
doi: 10.1023/B:SOLA.0000033366.10725.a2, 2004.

\bibitem[Chen et al.(2012)]{Chen2012}
Chen, G., Xu, J., Wang, W., Lei, J., Burns, A.G.,
A comparison of the effects if CIR- and CME--induced geomagnetic activity on thermospheric densities and spacecraft orbits: Case studies.
J. Geophys. Res. Space Physics, 117, A08315, doi: 10.1029/2012JA017782, 2012.

\bibitem[Chen et al.(2014)]{Chen2014}
Chen, G., Xu, J., Wang, W., Burns, A.G.,
A comparison of the effects if CIR- and CME--induced geomagnetic activity on thermospheric densities and spacecraft orbits: Statistical studies.
J. Geophys. Res. Space Physics, 119, 7928--7939, doi: 10.1002/2014JA019831, 2014.

\bibitem[Cook(1965)]{Cook1965}
Cook, G.,
Satellite drag coefficients. Planetary and Space Science, 13, 929-946, 1965.

\bibitem[Davis and Sugiura(1966)]{DavisSugiura1966}
Davis, T.N., and Sugiura, M.,
Auroral electroject activity index AE and its universal time variations.
J Geophys Res., 71, 785?801, 1966.

\bibitem[Doornbos et al.(2009)]{Doornbos2009}
Doornbos, E.,F\"orster, M.,Fritsche, B.,Helleputte, T.V.,Ijssel, J.V.D.,Koppenwallner, W., L\"uhr, H., Rees, D., Visser, P.,
Air density models derived from multi-satellite drag observations,Technical Report,DEOS TU Delft,317, 2009.

\bibitem[Gopalswamy(2000)]{Gopalswamy2000}
Gopalswamy, N., Lara, A., Lepping, R.~P., Kaiser, M.~L., Berdichevsky, D., St.~Cyr, O.~C.,
Interplanetary acceleration of coronal mass ejections.
Geophys. Res. Lett., 27, 145-148, doi: 10.1029/1999GL003639, 2000.

\bibitem[Gopalswamy(2006)]{Gopalswamy2006}
Gopalswamy, N.,
Properties of Interplanetary Coronal Mass Ejections, Space Science Reviews, 124, 145-168, doi: 10.1007/s11214-006-9102-1, 2006

\bibitem[Gosling et al.(1991)]{Gosling1991}
Gosling, J.T., McComas, D.J., Phillips, J.L., Bame, S.J.,
Geomagnetic activity associated with Earth passage of interplanetary shock disturbances and coronal mass ejections.
J. Geophys. Res. Space Physics, 96, A5, 7831--7839, doi: 10.1029/91JA00316, 1991.

\bibitem[Guo et al.(2010)]{Guo2010}
Guo, J., Feng, X., Forbes, J.M., Lei, J., Zhang, J.,
On the relationship between thermosphere density and solar wind parameters during intense geomagnetic storms.
J. Geophys. Res., 115, A12335, 2010.

\bibitem[Illing and Hundhausen(1985)]{Illing1985}
Illing, R.M.E., Hundhausen, A.J.,
Observation of a coronal transient from 1.2 to 6 solar radii.
J. Geophys. Res. 90: doi: 10.1029/JA090iA01p00275, 1985.

\bibitem[Iyemori et al.(2010)]{Iyemori2010}
Iyemori, T., Takeda, M., Nose M., Odagi Y., Toh, H.,
Mid-latitude Geomagnetic Indices ``ASY'' and ``SYM'' for 2009 (Provisional),
Data Anal. Cent. for Geomagn. and Space Magn., Faculty of Sci., Kyoto Univ., Kyoto, Japan.

\bibitem[Kamide and Akasofu(1983)]{Kamide1983}
Kamide, Y., Akasofu, S.I.,
Notes on the auroral electrojet indices,
Reviews of Geophysics and Space Physics, 21, 7, 1647-1656, doi: , 1983.

\bibitem[Kilpua et al.(2013)]{Kilpua2013}
Kilpua, E.K.J., Isavnin, A., Vourlidas, A., Koskinen, H.E.J., Rodriguez, L.,
On the relationship between interplanetary coronal mass ejections and magnetic clouds. Ann. Geophys., 31, 1251-1265, 2013.

\bibitem[King and Papitashvili(2004)]{King2004}
King, J.H.,Papitashvili, N.E.,
Solar wind spatial scales in and comparisons of hourly Wind and ACE plasma and magnetic field data,
J. Geophys. Res., Vol. 110, No. A2, A02209, 10.1029/2004JA010804, 2004.

\bibitem[Knipp et al.(2004)]{Knipp2004}
Knipp, D.~J., Tobiska, W.~K., Emery, B.~A., Direct and Indirect Thermospheric Heating Sources for Solar Cycles 21-23.
Solar Physics, 224, 495-505, doi: 10.1007/s11207-005-6393-4, 2004.

\bibitem[Krauss et al.(2012)]{Krauss2012}
Krauss, S., Fichtinger B., Lammer, H., Hausleitner, W., Kulikov, Y.N., Ribas, I., Shematovich, V., Bisikalo, D., Lichtnegger, H.I.M.,  Zaqarashvili, T.V., Khodachenko, M.L., Hanslmeier, A.,
Solar flares as proxy for young Sun: satellite observed thermosphere response to an X17.2 flare of Earth's upper atmosphere.
Ann. Geophys., 30, 1129-1141, doi: 10.5194 \textbackslash angeo-30-1129-2012, 2012.

\bibitem[Krauss(2013)]{Krauss2013}
Krauss, S.,
Response of the Earth's thermosphere during extreme solar events. A contribution of satellite observations to atmospheric evolution studies.
PhD. thesis, Graz University of Technology, 2013.

\bibitem[Liu et al.(2010)]{Liu2010}
Liu, R., L{\"u}hr, H., Ma, S.Y.,
Storm-time related mass density anomalies in the polar cap as observed by CHAMP.
Ann. Geophys., 28, 165-180, 2010.

\bibitem[Liu et al.(2014)]{Liu2014}
Liu, Y.D., Yang, Z., Wang, R., Luhmann, J.G., Richardson, J.D., Lugaz, N.,
Sun-to-Earth Characteristics of Two Coronal Mass Ejections Interacting near 1 AU: Formation of a Complex Ejecta and Generation of a Two-Step Geomagnetic Storm.
The Astrophysical Journal Letters, 793:L41, doi: 10.1088/2041-8205/793/2/L41, 2014.

\bibitem[Lukianova et al.(2002)]{Lukianova2002}
Lukianova, R., Troshichev O.,
The polar cap magnetic activity indices in the southern (PCS) and northern (PCN) polar caps: Consitency and discrepancy,
Geophys. Res. Lett., 29, 18, 1879, doi: 10.1029/2002GL015179, 2002.

\bibitem[Mayaud(1968)]{Mayaud1968}
Mayaud, P.N.,
Indices Kn, Ks, Km, 1964?1967,
Centre National de la Recherche Scientifique, Paris, 156p, 1968.

\bibitem[Mayaud(1971)]{Mayaud1971}
Mayaud, P.N.,
Une mesure plan\'etaire d'activit\'e magn\'etique bas\'ee sur deux observatoires antipodaux.
Ann. Geophys., 27, 67, 1971.

\bibitem[McComas et al.(1998)]{McComas1998}
McComas, D.J., Bame, S.J., Barker, P., Feldman, W.C., Phillips, J.L., Riley, P., Griffee, J.W.,
Solar Wind Electron Proton Alpha Monitor (SWEPAM) for the Advanced Composition Explorer. Space Science. Rev., 86, 563?612, 1998.

\bibitem[Menvielle et al.(2011)]{Menvielle2011}
Menvielle, M., Iyemori, T., Marchaudon, A., Nos\'{e}, M.
Geomagnetic indices in Geomagnetic Observations and Models.
Edited by M. Mandea nad M. Korte, IAGA Special Sopron Book Series, Vol 5., Springer+Business Media 183-228, ISBN: 978-90-481-9857-3, doi 10.1007/978-90-481-9858-0, 2011.

\bibitem[Reigber et al.(2002)]{Reigber2002}
Reigber, C., L{\"u}hr, H., Schwintzer P.,
CHAMP mission status. Advances in Space Research, 30(2), 129-134, 2002.

\bibitem[Richardson and Cane(2003)]{Richardson2003}
Richardson, I.G., Cane, H.V.,
Interplanetary coronal mass ejections in the near-Earth solar wind during 1996-2002. J. Geophys. Res., 108,A4,1156, doi:10.1029/2002JA009817, 2003.

\bibitem[Richardson and Cane(2010)]{Richardson2010}
Richardson, I.G., Cane, H.V.,
Near-Earth Interplanetary Coronal Mass Ejections During Solar Cycle 23 (1996--2009): Catalog and Summary of Properties.
Solar Physics,264,1, 189-237,2010.

\bibitem[Sentman(1961)]{Sentman1961}
Sentman, L., Free molecule flow theory and its application to the determination of aerodynamic forces, Technical Report, 1961.

\bibitem[Smith et al.(1998)]{Smith1998}
Smith, C., L'Heureux, J., Ness, N., Acuna, M., Burlaga, L., Scheifele, J.,
The ACE Magnetic Fields Experiment,
Space Science, Rev., 86, 613?632, doi:10.1023/A:1005092216668, 1998.

\bibitem[St.\ Cyr et al.(2000)]{StCyr2000}
St. Cyr, O.C., Howard, R.A., Sheeley, N.R., Plunkett, S.P., Michels, D.J., Paswaters, S.E., Koomen, M.J., Simnett, G.M., Thompson, B.J., Gurman, J.B., Schwenn, R.,Webb, D.F., Hildner,
E., Lamy, P.L.,
Properties of coronal mass ejections: SOHO LASCO observations from January 1996 to June 1998,
J. Geophys. Res., 105, 18169?18186, doi:10.1029/1999JA000381, 2000.

\bibitem[Stone et al.(1998)]{Stone1998}
Stone, E.C., Frandsen, A.M., Mewaldt, R.A., Christian E.R., Margolies, D., Ormes, J.F.,Snow, F.,
The Advanced Composition Explorer,
Space Science Reviews, 86, 1-4, pp. 1--22, doi: 10.1023/A:1005082526237, 1998.

\bibitem[Sugiura(1964)]{Sugiura1964}
Sugiura, M.
Hourly values of equatorial Dst for the IGY.,
Annals of International Geophysics Year, 35, 9,. Pergamon Press, Oxford, 1964.

\bibitem[Sugiura and Kamei(1991)]{SugiuraKamei1991}
Sugiura, M. and Kamei, T.,
Equatorial Dst index 1957--1986
IAGA Bulletin No. 40, 1991.

\bibitem[Sutton(2008)]{Sutton2008}
Sutton, E.K.,
Effects of Solar Disturbances on the Thermosphere Densities and Winds from CHAMP and GRACE Satellite Accelerometer Data.
PhD. thesis, Department of Aerospace Engineering Sciences, University of Colorado, 2008.

\bibitem[Tapley et al.(2004)]{Tapley2004}
Tapley, B.D., Bettadpur, S., Watkins, M., Reigber, C.,
The Gravity Recovery and Climate Experiment: Mission Overview and Early Results,
Geophysical Research Letters, 31, 9, doi:10.1029/2004GL019920, 2004.

\bibitem[Temmer et al.(2008)]{Temmer2008}
Temmer, M., Veronig, A.M., Vr\v{s}nak, B., Ryb\'{a}k, J., G\"om\"ory, P., Stoiser, S., Mari\v{c}i\'{c}, D.,
Acceleration in Fast Halo CMEs and Synchronized Flare HXR Bursts.
The Astrophysical Journal Letters, 673, L95, doi:10.1086/527414, 2008.

\bibitem[Temmer and Nitta(2015)]{Temmer2015}
Temmer, M., Nitta, N. V.,
Interplanetary Propagation Behavior of the Fast Coronal Mass Ejection on 23 July 2012,
Solar Physics, 290, 919--932, doi: 10.1007/s11207-014-0642-3, 2015.

\bibitem[Thayer et al.(2012)]{Thayer2012}
Thayer, J.P., Liu, X., Lei, J., Pilinski, M., Burns, A.G.,
The impact of helium on the thermosphere mass density response to geomagnetic activity during the recent solar minimum.
J. Geophys. Res., 117, A07315, doi:10.1029/2012JA017832, 2012.

\bibitem[Troshichev et al.(1988)]{Troshichev1988}
Troshichev, O.A., Andrezen, V.G., Vennerstr\o m, S., Friis--Christensen, E.,
Magnetic activity in the polar cap -- A new index.
Planet Space Science, 36, 1095, 1988.

\bibitem[Vassiliadis et al.(1996)]{Vassiliadis1996}
Vassiliadis, D., Angelopoulos, V., Baker, D.N., Klimas, A.J.,
The relation between the northern polar cap and auroral electrojet geomagnetic indices in the wintertime.
Geophysical Research Letters, 23, 20, 2781--2784, 1996.

\bibitem[Vasyli\={u}nas and Song(2005)]{Vasyliunas2005}
Vasyli\={u}nas, V.M., Song, P.,
Meaning of ionospheric joule heating.
J. Geophys. Res., 110, A02301, doi: 10.1029/2004JA010615, 2005.

\bibitem[Vourlidas et al.(2013)]{Vourlidas2013}
Vourlidas, A., Lynch, B.J., Howard, R.A., Li, Y.
How Many CMEs Have Flux Ropes? Deciphering the Signatures of Shocks, Flux Ropes, and Prominences in Coronagraph Observations of CMEs.
Solar Physics, 284,1,179-201,doi: 10.1007/s11207-012-0084-8, 2013.

\bibitem[Vr\v{s}nak(2008)]{Vrsnak2008}
Vr{\v s}nak, B., Processes and mechanisms governing the initiation and propagation of CMEs.
Ann. Geophys., 26, 3089-3101,
doi:10.5194/angeo-26-3089-2008, 2008.

\bibitem[Vr\v{s}nak et al.(2007)]{Vrsnak2007}
Vr\v{s}nak, B., Mari{\v c}i{\'c}, D., Stanger, A.~L., Veronig, A.~M., Temmer, M., Ro{\v s}a, D.,
Acceleration Phase of Coronal Mass Ejections: I. Temporal and Spatial Scales. Solar Physics, 241, 85-98, doi:10.1007/s11207-006-0290-3, 2007.

\bibitem[Vr\v{s}nak et al.(2013)]{Vrsnak2013}
Vr{\v s}nak, B., \v{Z}ic, T., Vrbanec, D., Temmer, M., Rollett, T., M{\"o}stl, C., Veronig, A., {\v C}alogovi{\'c}, J.,
Dumbovi{\'c}, M., Luli{\'c}, S., Moon, Y.-J., Shanmugaraju, A., Propagation of Interplanetary Coronal Mass Ejections: The Drag-Based Model.
Solar Physics, 285, 295-315, doi: 10.1007/s11207-012-0035-4, 2013.

\bibitem[Wanliss and Showalter(2006)]{Wanliss2006}
Wanliss, J.A., Showalter, K.M.,
High-resolution global storm index: Dst versus SYM-H.
J. Geophys. Res., 111, A02202, doi:10.1029/2005JA011034, 2006.

\bibitem[Wilson et al.(2006)]{Wilson2006}
Wilson G.R., Weimer R., Wise J.O., Marcos F.A.,
Response of the thermosphere to Joule heating and particle precipitation.
J. Geophys. Res., 111, A10314, doi:10.1029/2005JA011274, 2006.

\bibitem[Yue and Zong(2011)]{Yue2011}
Yue, C., Qiugang Zong, Q.,
Solar wind parameters and geomagnetic indices for four different interplanetary shock/ICME structures.
J. Geophys. Res., 116, A12, doi:10.1029/2011JA017013, 2011.

\bibitem[Yurchyshyn et al.(2005)]{Yurchyshyn2005}
Yurchyshyn, V., Yashiro, S., Abramenko, V., Wang, H., Gopalswamy, N.,
Statistical distributions of speeds of coronal mass ejections.
The Astrophysical Journal, 619, 599?603, doi:10.1086/426129, 2005.

\bibitem[Zhang et al.(2001)]{Zhang2001}
Zhang, J., Dere, K.~P., Howard, R.~A., Kundu, M.~R., White, S.~M., On the Temporal Relationship between Coronal Mass Ejections and Flares.
The Astrophysical Journal Letters, 559, 452-462, doi:10.1086/322405, 2001.

\end{thebibliography}
\end{document}